\documentclass[preprint,12pt]{elsarticle}

\graphicspath{ {./images/} }
\usepackage{subfigure}

\usepackage{amssymb}

\journal{The name of international journal CS}

\begin{document}

\begin{frontmatter}



\title{The Multi-granularity in Graph Revealed by a Generalized Leading Tree}


\author{Shun Fu}
\author{}
\author{Ji Xu}

\address{}

\begin{abstract}
There are hierarchical characteristics in the network and how to effectively reveal the hierarchical characteristics in the network is a problem in the research of network structure. If a node is assigned to the community to which it belongs, how to assign the community to a higher level of community to which it belongs is a problem. In this paper, the density of data points is investigated based on the clustering task. By forming the density of data points, the hierarchical difference of data points is constructed. In combination with the distance between data points, a density-based leading tree can be constructed. But in a graph structure, it is a problem to build a lead tree that reveals the hierarchical relationships of the nodes on the graph. Based on the method of tree formation based on density, this paper extends the model of leading tree to the hierarchical structure of graph nodes, discusses the importance of graph nodes, and forms a leading tree that can reveal the hierarchical structure of graph nodes and the dependency of community. Experiments were carried out on real data sets, and a tree structure was formed in the experiment. This graph leading tree can well reveal the hierarchical relationships in the graph structure.


\end{abstract}

\begin{keyword}
Granular computing \sep Network representation learning \sep Graph clustering


\end{keyword}

\end{frontmatter}


\section{Introduction}
\label{sec.intro}

Graph is widely used to model and display the relationships between objects and objects in the world. On the other hand, the distribution of many data objects presents multi-level characteristics, such as the organization and management structure of human beings, the food chain of the earth, the taxonomy of species, and so on. In the relational network (such as social network, academic network, traffic network, etc.), the distribution of nodes also presents multi-level characteristics. Simply taking a single node as an object to study, the problem will be very complicated. On the other hand, people can study the objects in the figure at different levels by referring to the cognitive rules and multi-granularity cognitive methods of human beings for complex problems. [question] : which nodes in the network are at the higher level and which are at the lower level? Which nodes make up the community? And which communities form a larger community? Answering these questions requires looking at complex relational networks from a multi-granularity perspective, in which objects (nodes, edges, communities, etc.) can be subordinate to different granularity levels. Therefore, it is important to reveal the multi-granularity and multi-level structure of node objects in the network.


\subsection{The hierarchical properties in networks}
(introduction and related works)  

\subsection{Hierarchical clustering}
(introduction and related works)

\begin{figure}[h]
\centering
  \subfigure[caption1]{
  \label{fig1.sub1}
  \includegraphics[width=0.3 \linewidth]{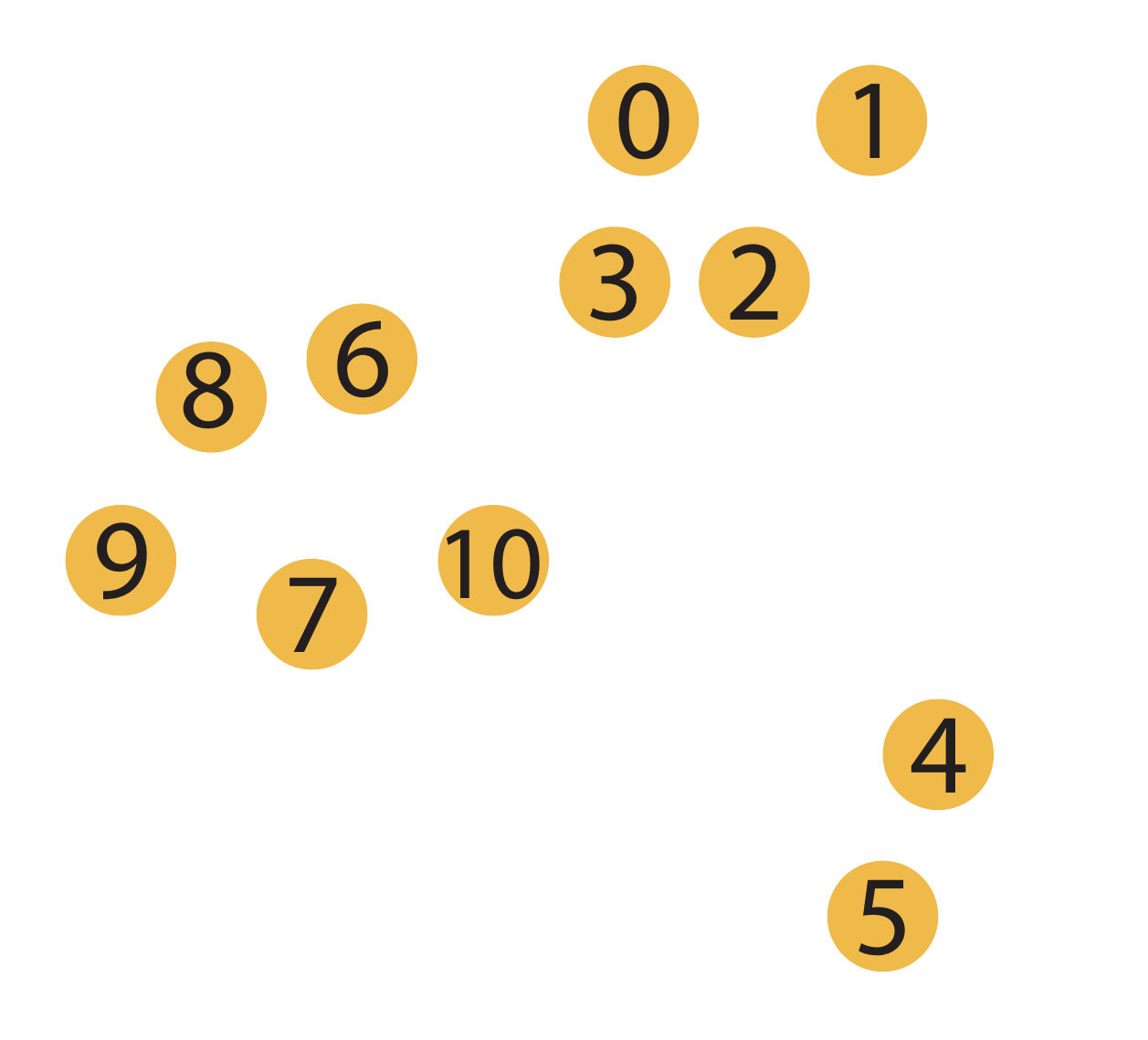}
}
  \subfigure[caption2]{
  \label{fig1.sub2}
  \includegraphics[width=0.3 \linewidth]{sec101.pdf}
}
  \subfigure[caption3]{
  \label{fig1.sub3}
  \includegraphics[width=0.3 \linewidth]{sec101.pdf}
}

  \caption{Caption for subfigures}
  \label{fig.str_concern}
\end{figure}

\subsection{The leading tree for clustering in Euclidean space}

When performing efficient and accurate clustering of general data points in datasets of any shape in European space, Rodriguez and Laio proposed the DPClust algorithm \cite{rodriguez2014clustering}. Xu \cite{xu_DenPEHC_2016} Considered the cluster assignment process of the DPClust algorithm as a process of finding its parent node for each data point. Once the search for the parent node is completed, a tree-like subordinate relationship is formed between the nodes. As shown in Figure \ref{fig1.sub1} in the European space, the tree structure formed by the DPClust algorithm. The application of Leading Tree in the field of hierarchical clustering makes it easy to examine the multi-level community relationships in the distribution of data points at multiple granularities. As shown in Figure \ref{fig1.sub2} Leading Tree maps data to a tree-shaped space. One can easily see that the points in the upper layer of the tree are more at the core position of the cluster, while the points in the lower layer correspond to the edge points of a cluster. In other words, the mapping to the tree structure assigns points in the original European space to Granular Layers. On the other hand, this tree structure can be broken down into many subtrees, and small subtrees can be merged into large subtrees. This corresponds to the multi-level division of clusters in the original Euclidean space. As shown in Figure \ref{fig1.sub3}, subtree a corresponds to class cluster A, and a belongs to a larger subtree b, so in a coarser granularity, class cluster A is a component of class cluster B.

For the points of Euclidean space, DenPEHC \cite{xu_DenPEHC_2016} has realized the rapid and effective multi-level clustering and the clear, intuitive and effective disclosure of the multi-level dependency of these points through the construction of DenPEHC guidance tree based on density. But for the data objects in our widely used Graph space, how can we construct a multi-grained tree structure to reveal the above relationships?


The construction process of Leading Tree in DenPEHC is carried out for data points in Euclidean space, which relies on the distance between data points and the artificially defined density given in data points. So that the point is subordinate to the point that is closest to it, that is denser. We believe that in Euclidean space, the process of a data point finding its parent node can be more generally extended to data objects in any space, such as graph data objects. The purpose of this paper is to discuss how to extend the idea of leading tree construction to the data object of graph space so that this idea of multi-granularity computation can reveal the hierarchical relationship of data distribution in graph space.


\section{Contribution}
\label{sec.contributions}
In this paper, we offer contributions by :
1. Proposing a novel particle calculation method is introduced to reveal the structure hierarchy of graph nodes
2. Provide a new tool and method for Graph Clustering, which can effectively and intuitively reveal the structure hierarchy of Graph data objects
3. The leading tree only applicable to European type space is expressed in a more general way, and it is extended to graph data objects.
4. The measurement standard of the importance of graph nodes is discussed to inspire the exploration of this problem.

\section{Related works}
\label{sec.rel_works}

\subsection{Granular computing}
Granular computing is an umbrela concept which aims revealing the hierarchical relationships between objects \cite{yao2013granular}. Granule is the basic unit of object for processing in granular computing. The granule can be defined as different types and they are flexible and scalable. ...

\subsection{Hierarchical Clustering}

\subsection{Graph Clustering}

\section{Problems}
\label{sec.problem}

\subsection{A generalized model of leading tree}
For any object, or granule, you can define its importance and its distance from other granule of the same kind. With these two elements, you can generate a tree structure. This tree structure will reveal the membership and hierarchy of all granule classes in a set. This tree structure will form a general lead tree. Not limited to data points in Euclidean space.

For different data objects, or granule, its importance needs to be either manually defined or machine-learned depending on the needs of the problem. The distance between two objects also needs to be defined according to the needs of the problem.

\begin{figure}[h]
    \includegraphics[width=0.99 \linewidth]{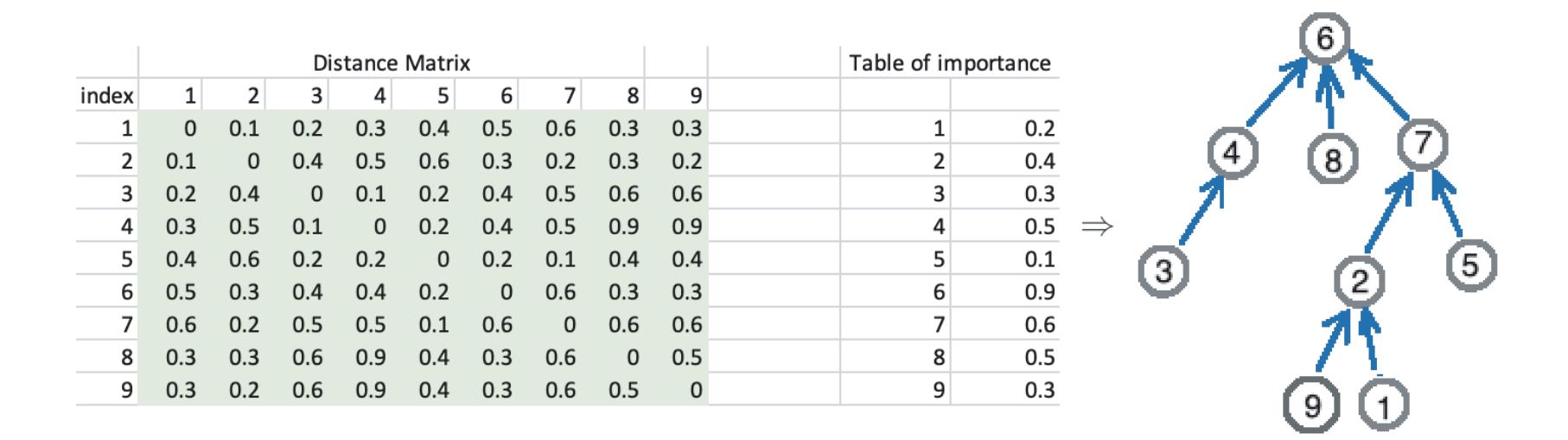}
    \caption{Two elements for GLT generation: tabel of importance and matrix of distance}
    \label{fig.gen_tree}
\end{figure}

Take a set of 9 granule as an example. The distance matrix between them is shown in the Figure \ref{fig.gen_tree}. Based on the distance matrix and the importance table, the parent of a granule can be found. This generates the tree structure on the right side of the diagram.

With the tree structure, the dependencies between granule are clearly visible. Here, a granule can be a point in a Euclidean space, a vertex in the Graph, a person in a collaborating organization, a community of vertices, and so on.

The tree structure in Fig.\ref{fig.gen_tree} reflects the membership of the nine granule. 1,9 is led by 2. One, two, nine constitute a small grass-roots organization. Then they are led by 7, along with 5. So {1,2,9,5,7} constitutes a higher-level organization centered on 7. These organizational levels are abstract concepts, and if granule is materialized, the meaning of these levels can be materialized, such as the community membership formed by nodes in the graph, the management structure of a company, or the wildlife food chain.



\section{Distance and importance measuring in Graph}
\label{measure.graph}
As we saw in the previous section, the necessary conditions for building a generalized granule leading tree are the distance (similarity) measure between granule and the importance of granule under certain criteria. In this section we extend the Generalized Leading Tree defined in the previous section specifically to the Graph problem. Firstly, in the data point set of Euclidean space, the distance measure can be defined by the cosine distance between the data point vectors. If the logarithmic data points are clustered, the importance measure can be the density of a certain point.


In order to extend the general Tree model discussed in the previous section to the facilities of revealing the  Hierarchical/multi - granularity of Property in Graph, we need to define the distance between the granules and the importance measurment for granules in Graph. This section lists several measures and discusses the information focused on under different measures. In the experimental section, we will reveal the differences in the lead trees generated under different metrics and discuss the hierarchical nature of the information expressed through different lead trees.


\subsection{The importance measure by degrees}  
The importance of a node can be measured by its degree. Degree is the most directe sign which counts the link between the vertex to other vertices. In most senarios, degree reflects the influence capability of one vertex. For example, in a community, if a vertex has the largest value of degree, there should be some unique properties in that vertex. Maybe that vertex is the most social active member in that community.

\subsection{The importance measure by eigenvector centrality} 

Eigenvector centrality is a kind of extension of degree—it looks at a combination of a node’s edges and the edges of that node’s neighbors. Eigenvector centrality cares if you are a hub, but it also cares how many hubs you are connected to. It’s calculated as a value from 0 to 1: the closer to one, the greater the centrality. Eigenvector centrality is useful for understanding which nodes can get information to many other nodes quickly. If you know a lot of well-connected people, you could spread a message very efficiently. If you’ve used Google, then you’re already somewhat familiar with Eigenvector centrality. Their PageRank algorithm uses an extension of this formula to decide which webpages get to the top of its search results.

\subsection{The importance measure by betweenness centrality} 

Betweenness centrality is a bit different from the other two measures in that it doesn’t care about the number of edges any one node or set of nodes has. Betweenness centrality looks at all the shortest paths that pass through a particular node (see above). To do this, it must first calculate every possible shortest path in your network, so keep in mind that betweenness centrality will take longer to calculate than other centrality measures (but it won’t be an issue in a dataset of this size). Betweenness centrality, which is also expressed on a scale of 0 to 1, is fairly good at finding nodes that connect two otherwise disparate parts of a network. If you’re the only thing connecting two clusters, every communication between those clusters has to pass through you. In contrast to a hub, this sort of node is often referred to as a broker. 

\subsection{The importance measure by the length of path} 

In addition to the measure of importance of vertex, we need to measure the distance bwteen vertices in graphs for the GLT generation. The most simple way is the length of path between two vertices. Another way is using the similarity measurement and minused by one to obtain the distance between vertices. The similarity measurement can be Jaccard coefficient, SimRank, PageRank etc.

\section{The graph leading tree model}  
\label{}

For vertex $v$, in the set of vertices that has higher degree than $v$, we take the closest one (e.g. $u$) as the parent vertex of $v$. By analogy, one or more trees are formed to form an interpretable semantic tree or semantic forest that reveals hierarchical semantic concepts on the structure of the relational network.



\bibliographystyle{elsarticle-num} 
\bibliography{myBibDataBase}






\end{document}